\documentclass[12pt]{article}
\usepackage{a4wide}
\usepackage{amssymb}
\usepackage{amsmath}
\usepackage{graphicx}
\usepackage{mathdots}
\usepackage{mathrsfs}

\def\makeatletter{\catcode`\@=11}
\makeatletter
\def\mathbox#1{\hbox{$\m@th#1$}}%
\def\math@ccstyles#1#2#3#4#5#6#7{{\leavevmode
      \setbox0\mathbox{#6#7}%
      \setbox2\mathbox{#4#5}%
      \dimen@ #3%
      \baselineskip\z@\lineskiplimit#1\lineskip\z@
      \vbox{\ialign{##\crcr
             \hfil \kern #2\box2 \hfil\crcr
             \noalign{\kern\dimen@}%
             \hfil\box0\hfil\crcr}}}}
\def\mathaccstyles{\math@ccstyles\maxdimen}
\def\maththroughstyles{\math@ccstyles{-\maxdimen}}
\def\unity%
 {\maththroughstyles{.45\ht0}\z@\displaystyle {\mathchar"006C}\displaystyle 1}

\begin{document}

\setcounter{table}{0}

\mbox{}
\vspace{2truecm}
\linespread{1.1}

\centerline{\LARGE \bf Probing the Higgs branch of 5d fixed point theories}

\vspace{.5cm}

 \centerline{\LARGE \bf with dual giant gravitons in $AdS_6$}

\vspace{2truecm}

\centerline{
    {\large \bf Oren Bergman ${}^{a}$} \footnote{bergman@physics.technion.ac.il}
     {\bf and}
    {\large \bf Diego Rodr\'{\i}guez-G\'omez${}^{a}$} \footnote{drodrigu@physics.technion.ac.il}}

\vspace{1cm}
\centerline{{\it ${}^a$ Department of Physics, Technion, Israel Institute of Technology}} \centerline{{\it Haifa, 32000, Israel}}
\vspace{1cm}

\centerline{\bf ABSTRACT}
\vspace{1truecm}

\noindent

We consider the warped $AdS_6\times S^4/\mathbb{Z}_n$ backgrounds dual to certain 5d quiver gauge theories. 
By studying dual giant gravitons in the $AdS_6$ geometry we are able to partially probe the 
Higgs branch of these theories. 
We show how the quantization of the phase space of such dual giants coincides with the counting of 
holomorphic functions on $\mathbb{C}^2/\mathbb{Z}_n$, which is the geometric part of the Higgs branch for these theories.

\newpage

\tableofcontents

\section{Introduction}

Generically, 5d gauge theories do not exist as microscopic theories since they are non-renormalizable and thus require a UV completion beyond the scale set by the inverse Yang-Mills coupling. However,
under certain circumstances, it is possible to remove the UV cutoff while having a theory well defined everywhere on its moduli space
\cite{Seiberg:1996bd, Morrison:1996xf,Intriligator:1997pq}. 
The crucial point is that minimally supersymmetric theories in 5d contain 8 supercharges and therefore a non-abelian $SU(2)_R$ R-symmetry. The effective action on the Coulomb branch follows then from a pre-potential, which in the 5d case is severely restricted by gauge-invariance and anomaly considerations.\footnote{In 5d, upon integration out  massive fermions, a Chern-Simons term is produced \cite{Witten:1996qb}. This is very similar to the 3d parity anomaly.} Inspection of the exact effective gauge coupling 
shows that, upon appropriately choosing the gauge group and matter content, the bare Yang-Mills coupling 
can be removed. The resulting theory 
is expected to be at an isolated fixed point.

A particularly interesting theory is that of a $USp(2\,N)$ gauge group with an antisymmetric hyper-multiplet and $N_f$ fundamental hyper-multiplets. According to the analysis in  \cite{Seiberg:1996bd, Morrison:1996xf,Intriligator:1997pq} this theory is at a fixed point as long as $N_f<8$. Moreover it can be naturally embedded into string theory as the world-volume theory of $N$ D4 branes probing an $O8^-$ plane with $N_f$ D8 branes on top of it \cite{Seiberg:1996bd}. From the string theory perspective, the inverse bare YM coupling corresponds to the value of the dilaton at the orientifold plane. The fixed point theory corresponds to the case where the dilaton is tuned to diverge on top of the O8/D8. The $SO(2\,N_f)$ global flavor symmetry, corresponding to the D8-brane gauge
symmetry, is then enhanced to $E_{N_f+1}$ via massless D0-brane states (dual to instanton particles in the 5d gauge theory) localized 
at the position of the orientifold \cite{Polchinski:1995df, Matalliotakis:1997qe, Bergman:1997py}. 
This has been recently 
demonstrated in \cite{Kim:2012gu} from a purely field theoretical perspective. 
The near-horizon limit of this brane construction gives a warped $AdS_6\times S^4$ background in
massive Type IIA supergravity \cite{Brandhuber:1999np}, reinforcing the claim that the 5d field theory under 
consideration is indeed at a fixed point.

Starting with this basic theory, three infinite families of daughter theories were constructed in \cite{Bergman:2012kr} by replacing 
the flat $\mathbb{R}^4$ transverse to the D4's inside the $O8/D8$ by an orbifold $\mathbb{C}^2/\mathbb{Z}_n$. This produces
quiver gauge theories involving products of $USp(2\,N)$ and $SU(2\,N)$ gauge groups,
with dual massive Type IIA supergravity backgrounds given by warped $AdS_6\times S^4/\mathbb{Z}_n$. 
The $S^5$ free energy of the quiver theories was recently computed using localization in 
\cite{Jafferis:2012iv}, and shown to agree precisely with the entanglement entropy for an $S^4$ in supergravity,
thus providing further support for the existence of the quiver fixed points and for the $AdS_6$ duals.  Note that supersymmetric $AdS_6$ solutions are remarkably hard to find \cite{Passias:2012vp}, thus rendering this series of examples is quite noteworthy. Interestingly, upon allowing for more exotic ansatze one can find other $AdS_6$ solutions \cite{AdS6}.

We expect that the quiver theories also exhibit 
an enhanced $E_{N_f+1}$ global symmetry on the Higgs branch. Note that, on general grounds (see \textit{e.g.} \cite{Tong:2005un}), the Higgs branch of these theories coincides with the moduli space of $E_{N_f+1}$ instantons on $\mathbb{C}^2/\mathbb{Z}_n$. This enhanced $E_{N_f+1}$ symmetry is not visible in the gravity dual. In fact, the latter becomes singular at the location of the O8/D8 as both the dilaton and the curvature diverge. This is not surprising since certainly supergravity fails to capture the D0-branes which become massless and provide the necessary extra states for the symmetry enhancement. Nevertheless, the full Higgs branch of the theory contains operators which are both flavor- and instanton-blind, and thus are insensitive to this symmetry enhancement. In this paper we concentrate on such operators, which can then be thought of as (partial) probes of the Higgs branch. In the gravity dual they correspond to dual giant gravitons sitting on top of the O8/D8. As we will see, although the curvature and dilaton diverge at that point, the world-volume theory on the dual giants is perfectly well behaved, and in fact matches the expected field theory results upon geometric quantization of their phase space as in  \cite{Mandal:2006tk,Martelli:2006vh} (see also \cite{Kinney:2005ej, minwalla}).

The plan for the rest of the paper is as follows. In section \ref{review} we provide a lightning review of the quiver theories under consideration and their gravity duals. In section \ref{geodesics} we study a family of massless geodesics in the geometry. These massless geodesics are followed by the dual giant gravitons, which we introduce in section \ref{duals}. We then perform the geometric quantization of their phase space in section \ref{symplecticquantization} and find that it is in one-to-one correspondence to that of a $\mathbb{C}^2/\mathbb{Z}_n$ which can be actually thought as the pre-near horizon ALE space. This result is matched with the field theory expectations in section \ref{fieldtheory}. Finally, we end in section \ref{conclusions} with some comments and future prospects.

\section{5d quiver theories and their $AdS_6$ duals}
\label{review}

Following \cite{Bergman:2012kr}, the class of 5d theories of interest can be engineered by considering in type $I'$ string theory $N$ 
D4-branes probing an O8-plane with $N_f$ coincident D8-branes wrapping an ALE space as follows 
(the boxed coordinates denote the ALE directions),
\begin{equation}
\begin{array}{l c |  c c c c c c c c c}
& 0 & 1 & 2 & 3 & 4 &\boxed{5}& \boxed{6} &\boxed{7} &\boxed{8}  & 9 \\ \hline
D8/O8^- & \times & \times & \times & \times & \times & \times & \times & \times & \times &\\
D4 & \times & \times & \times & \times & \times &   &  &  & &\\
\end{array}
\,.
\label{syst}
\end{equation}

We can construct the corresponding theories by starting with Type IIA string theory on $\mathbb{C}^2/\mathbb{Z}_n$
and then performing the orientifold projection $\Omega\,I_9$. Prior to the orientifold we find an $\mathcal{N}=(1,\,1)$ 6d SUGRA multiplet together with $(n-1)$ 6d vector multiplets coming from the $(n-1)$ twisted sectors of the orbifold. Upon orientifolding this theory, since the orientifold involves an inversion, the resulting theory lives in 5d. Furthermore, due to the combined action of the inversion and the $\Omega$, the $i$-th twisted sector is identified with the $(n-i)$-th one, so that out of the original $n-1$, only half of them survive the orientifold projection, each giving rise to a 5d vector multiplet and a 5d hyper-multipelt. Obviously for the case of an even orbifold the middle twisted sector is left unpaired and hence it must be treated with special care. It turns out that there are two ways of implementing the orientifold projection on it  \cite{Polchinski:1996ry}: in one, which goes under the name of no vector structure (NVS), one keeps a 5d hyper-multiplet; while in the other, which goes under the name of vector structure (VS), one keeps the vector multiplet. In addition, in the NVS case there is trapped $B_2$ flux on the 2-cycle corresponding to the middle twisted sector. 

\begin{figure}[h!]
\centering
\includegraphics[scale=.7]{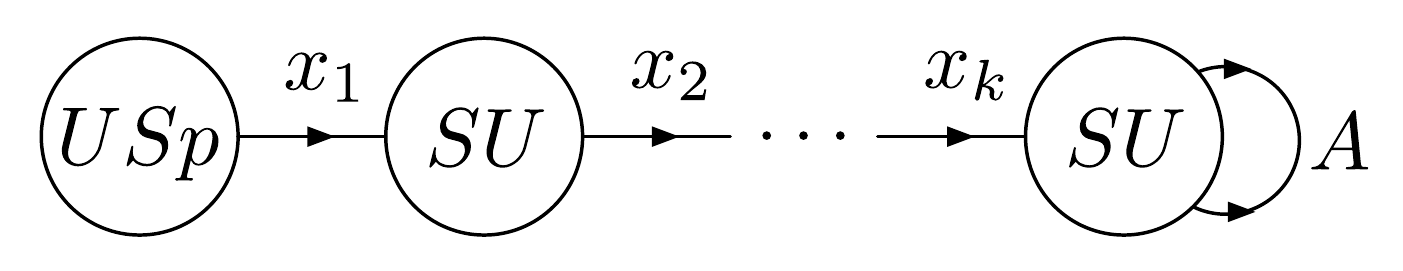} 
\caption{Quiver diagram for the $\mathbb{Z}_{2\,k+1}$ case.}
\label{Z2kplus1}
\end{figure}

\begin{figure}[h!]
\centering
\includegraphics[scale=.7]{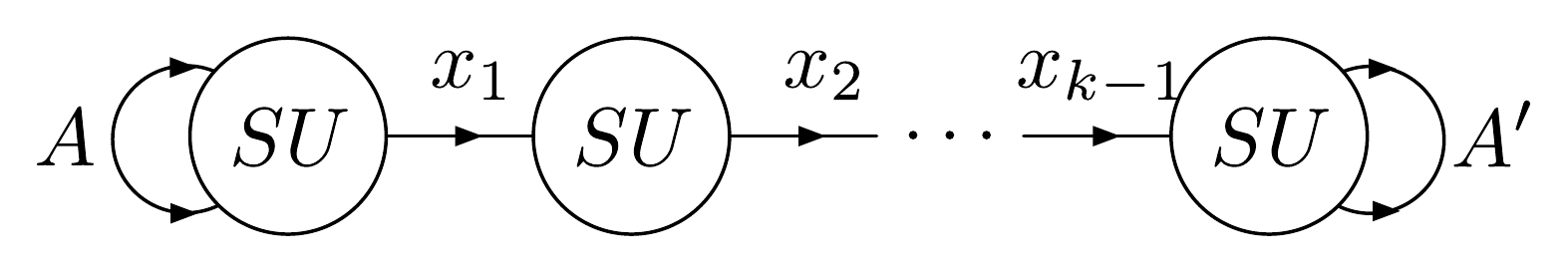} 
\caption{Quiver diagram for the $\mathbb{Z}_{2\,k}$ no vector structure case.}
\label{Z2knovectorstructure}
\end{figure}

\begin{figure}[h!]
\centering
\includegraphics[scale=.7]{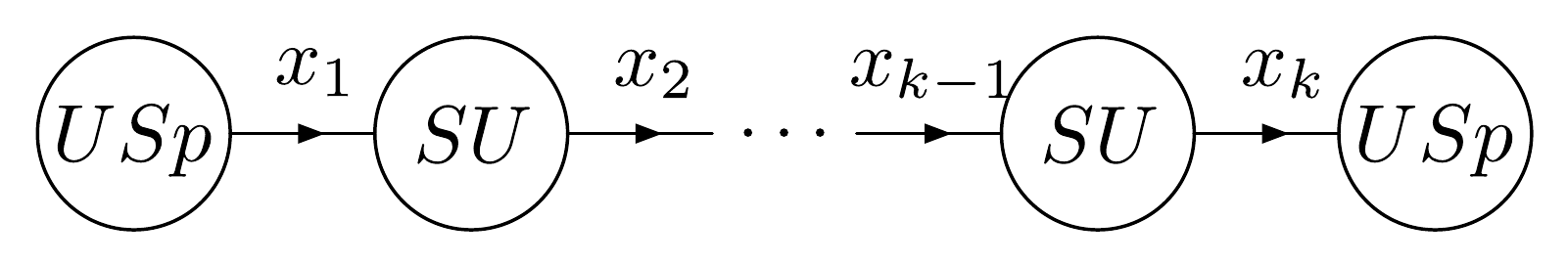} 
\caption{Quiver diagram for the $\mathbb{Z}_{2\,k}$ vector structure case.}
\label{Z2kvectorstructure2}
\end{figure}

The corresponding open string sectors must also be adjusted accordingly. 
The world-volume theories on the D4-branes depend crucially on the type of orbifold. Let us set $N_f=0$. For odd orbifolds $\mathbb{C}^2/\mathbb{Z}_{2\,k+1}$ we find a $USp(2\,N)\times SU(2\,N)^k$ gauge theory with bi-fundamentals and an antisymmetric 
hyper-multiplet for the last $SU$ group as shown in 
fig.~\ref{Z2kplus1}. Note that this theory has a $[U(1)^k]_B\times [U(1)^{k+1}]_I\times U(1)_M$ global non-R symmetry, where the subscripts $B$, $I$ and $M$ denote respectively baryonic, instantonic and mesonic symmetries.\footnote{In 5d gauge theories there is a topological current for each gauge group 
constructed out of its field strength as $j_I=\star(F\wedge F)$. 
Instantons, which in 5d are particle-like excitations, are electrically charged under these symmetries.}
For even orbifolds $\mathbb{C}^2/\mathbb{Z}_{2\,k}$ without vector structure the gauge group is $SU(2\,N)^k$ and the matter content
includes $k-1$ bi-fundamentals and two antisymmetric hyper-multiplets, 
as shown in fig.~\ref{Z2knovectorstructure}. The global symmetry group is in this case $[U(1)^k]_B\times [U(1)^k]_I\times U(1)_M$.
For even orbifolds $\mathbb{C}^2/\mathbb{Z}_{2\,k}$ with vector structure we have a $USp(2\,N)\times SU(2\,N)^{k-1}\times USp(2\,N)$ gauge theory with bi-fundamental matter, fig.~\ref{Z2kvectorstructure2}. In this case, the global symmetry group is $[U(1)^{k-1}]_B\times [U(1)^{k+1}]_I\times U(1)_M$.

The dual massive Type IIA supergravity backgrounds are warped $AdS_6\times S^4/\mathbb{Z}_n$ with a metric and dilaton given by
\begin{equation}
ds^2=\hat{\Omega}^2\,\Big\{ds^2_{AdS_6}+\frac{4}{9}\,L^2\big(d\alpha^2+\cos^2\alpha\,d\tilde{\Omega}_3^2\big) \Big\} \; , \;\;
e^{\Phi}=\frac{3}{2\,L}\,\Big(\frac{3}{2}\,m\,\sin\alpha \Big)^{-\frac{5}{6}}\,,
\end{equation}
where
\begin{equation}
\hat{\Omega}=\Big(\frac{3}{2}\,m\,\sin\alpha \Big)^{-\frac{1}{6}}\; , \;\; L^4=\frac{3^{8/3}\,\pi\,n\,N}{2^{2/3}\,m^{1/3}} \;, \;\;
m=\frac{8-N_f}{2\,\pi} \,,
\end{equation}
and $d\tilde{\Omega}_3^2$ stands for the metric of the lens space $S^3/\mathbb{Z}_n$,
\begin{equation}
d\tilde{\Omega}_3^2=\frac{1}{4}\,(d\psi-\cos\theta\,d\phi)^2+\frac{1}{4}\,(d\theta^2+\sin^2\,d\phi^2) \,,
\end{equation}
where $\psi\,\in\,[0,\,\frac{4\,\pi}{n}]$.
The background also includes a RR 4-form and 0-form,
\begin{equation}
F_0=m \;, \;\; 
\tilde{F}_4=\frac{10}{81}\,\Big(\frac{2}{3}\Big)^{\frac{2}{3}}\,
m^{\frac{1}{3}}\,L^4\,\sin^{\frac{1}{3}}\alpha\,\cos^3\alpha\,d\alpha\wedge d\psi\wedge \omega_2 \,,
\end{equation}
where $\omega_2=\sin\theta\,d\theta\wedge d\phi$.
Note that $\alpha\,\in\,[0,\,\frac{\pi}{2}]$, so the compact space is really a hemisphere with a boundary at $\alpha=0$.
We can interpret this as the result of the orientifold action which takes 
$\alpha\rightarrow -\alpha$.
Due to the $\alpha$-dependence of the warp factor, the background only exhibits the symmetry of the lens space, 
which is generically $SU(2)\times U(1)$. These symmetries correspond respectively to the $SU(2)_R$ and $U(1)_M$ in the field theory. 

The background is singular at $\alpha = 0$, where both the curvature and the dilaton diverge.
This makes some of the properties of this solution, like the on shell Euclidean action, ill-defined at the supergravity level.
This presumably requires a stringy resolution.
However, many properties remain well-defined, and are indeed consistent with the dual 5d gauge theories \cite{Bergman:2012kr}.
The dual giant gravitons that we will analyze below are also completely well-defined in this background.

\subsection{Global coordinates}

To analyze the dual giant gravitons it is convenient to work in global coordinates for $AdS_6$,
dual to radial quantization of the 5d CFT's.
The $AdS$ metric is then given by
\begin{equation}
ds_{AdS_6}^2=-(1+\frac{r^2}{L^2})\,dt^2+\frac{dr^2}{(1+\frac{r^2}{L^2})}+r^2\,d\Omega_4^2 \,.
\end{equation}
Dualizing the 4-form we get 
\begin{equation}
*\tilde{F}_4 = \tilde{F}_6=\frac{10}{3}\,r^4\,dt\wedge dr\wedge \omega_4 \,,
\end{equation}
where $\omega_4$ is the volume 4-form of the $S^4$ in the global $AdS_6$. Since there is no $H_3$ flux, and the possible 
$B_2$ flux can only be along internal directions we can integrate the 6-form to get the 5-form potential:
\begin{equation}
C_5=-\frac{2}{3}\,r^5\,dt\wedge \omega_4 \,.
\end{equation}

\section{A family of massless BPS geodesics}\label{geodesics}

There are two circles in the internal space 
$S^4/\mathbb{Z}_n$ on which we could naturally imagine particles orbiting, namely those parametrized by $\psi$ and $\phi$. Let us then consider a massless particle at fixed $\alpha,\,\theta$ moving along those coordinates. Note that since we will be interested in massless particles we need to use a Polyakov-like action obtained by introducing a world-line metric so that the zero mass limit is well-defined. More explicitly, denoting the world-line time by $\tau$, we consider $\{t(\tau),\,\psi(\tau),\,\phi(\tau)\}$. Upon gauge-fixing the world-line metric to one, the action reads
\begin{equation}
S=-\int \,d\tau\, \hat{\Omega}^2\,\Big[ (1+\frac{r^2}{L^2})\,\dot{t}^2-\frac{4\,L^2}{9\,n^2}\,\cos^2\alpha\,\Big( (\dot{\psi}+\frac{n}{2}\,\cos\theta\,\dot{\phi})^2+\frac{n^2}{4}\,\sin^2\theta\,\dot{\phi}^2\Big) \Big] \,,
\end{equation}
where the dot indicates a derivative with respect to the world-line coordinate $\tau$. We have rescaled $\psi$ so that it takes values in $[0,\,2\,\pi]$. 
The world-line  hamiltonian is 
\begin{equation}
H_{WL}=\frac{P_t^2}{4\,\hat{\Omega}^2\,(1+\frac{r^2}{L^2})}-\frac{9\,(4\,P_{\phi}^2+n^2\,P_{\psi}^2-4\,n\,P_{\phi}\,P_{\psi}\,\cos\theta)}{16\,L^2\,\hat{\Omega}^2\,\sin^2\theta\,\cos^2\alpha} \,.
\end{equation}
The constraint imposed by the world-line metric sets this to zero, which gives 
\begin{equation}
L\,\mathcal{H}=\frac{3}{2}\,\sqrt{1+\frac{r^2}{L^2}}\,\sqrt{4\,P_{\phi}^2+n^2\,P_{\psi}^2-4\,n\,P_{\phi}\,P_{\psi}\,\cos\theta}\,\frac{1}{\cos\alpha\,\sin\theta}\,,
\end{equation}
where $\mathcal{H} = P_t$ is the energy of the particle.
Clearly, the energy is minimized at $\alpha=0$. For $\theta$ there are two possible solutions:
\begin{equation}
\begin{array}{l}
{\rm a)} \cos\theta= \frac{n}{2}\,\frac{P_{\psi}}{P_{\phi}}\,\,\leadsto\,\, L\,\mathcal{H}=3\,P_{\phi}\,\sqrt{g^{AdS_6}_{tt}} \\ \\
 {\rm b})\,\cos\theta=\frac{2}{n}\,\frac{P_{\phi}}{P_{\psi}} \,\,\leadsto\,\, L\,\mathcal{H}=\frac{3\,n}{2}\,P_{\phi}\,\sqrt{g^{AdS_6}_{tt}} \,.
\end{array}
\end{equation}
Since $|\cos\theta|\leq 1$, it is clear that if $\frac{P_{\phi}}{P_{\psi}}\,>\,\frac{n}{2}$ the appropriate solution will be a), while if $\frac{P_{\phi}}{P_{\psi}}\,<\,\frac{n}{2}$ the appropriate solution will be b).

\section{Dual giant gravitons}\label{duals}

Now consider  a D4-brane wrapping $\{t,\,\Omega_4\}$, and assume that $\psi=\psi(t)$, $\phi=-\phi(t)$. 
The induced metric is given by (we again rescale $\psi$ so that it takes values in $[0,\,2\,\pi]$)
\begin{equation}
ds^2=\hat{\Omega}^2\,\Big\{-\Big((1+\frac{r^2}{L^2})-\frac{4\,L^2}{9\,n^2}\,\cos^2\alpha\,\Big[ (\dot{\psi}+\frac{n}{2}\,\cos\theta\,\dot{\phi})^2+\frac{n^2}{4}\,\sin^2\theta\,\dot{\phi}^2\Big]\Big)\,dt^2+r^2\,d\Omega_4^2\big) \Big\} \,.
\end{equation}
The D4-brane action is then given by
\begin{eqnarray}
\label{D4_action}
S&=&-\mu_4\,V_4\,\int \frac{2}{3}\,L\,r^4\,\sqrt{1+\frac{r^2}{L^2}}\sqrt{1-\frac{4\,L^2}{9\,n^2\,(1+\frac{r^2}{L^2})}\,\cos^2\alpha\,\Big[ (\dot{\psi}+\frac{n}{2}\,\cos\theta\,\dot{\phi})^2+\frac{n^2}{4}\,\sin^2\theta\,\dot{\phi}^2\Big]}\nonumber \\ && -\mu_4\,V_4\,\int\,\frac{2}{3}\,r^5 \,.
\end{eqnarray}
The equation of motion for $\alpha$ is again solved for $\alpha=0$.
Although this is a singular locus in the geometry, where both the curvature and the dilaton diverge, the behavior of BPS geodesics there is well-defined.

Legendre-transforming to the hamiltonian $\mathcal{H}=\mathcal{H}(P_{\psi},\,P_{\phi},\,\theta,\,r)$ we get
\begin{equation}
\mathcal{H}=\frac{3}{L}\,\sqrt{1+\frac{r^2}{L^2}}\,\sqrt{\frac{1}{\sin^2\theta}\,\Big( P_{\phi}^2+\frac{n^2}{4}\,P_{\psi}^2-n\,P_{\phi}\,P_{\psi}\,\cos\theta\Big)+\frac{4\,L^4\,\mu_4^2\,V_4^2}{81}\,r^8}-\frac{2}{3}\,\mu_4\,V_4\,r^5 \,.
\end{equation}
We again find two solutions for $\theta$ depending on the value of $P_\phi/P_\psi$:
\begin{equation}
\label{choices}
{\rm a})\,\cos\theta = \frac{n}{2}\,\frac{P_{\psi}}{P_{\phi}} \;\; \mbox{if} \;\; \frac{P_\phi}{P_\psi} > \frac{n}{2}
\qquad {\rm b})\,\cos\theta=\frac{2}{n}\,\frac{P_{\phi}}{P_{\psi}} \;\; \mbox{if} \;\; \frac{P_\phi}{P_\psi} < \frac{n}{2} \,.
\end{equation}
Plugging these solutions back into $\mathcal{H}$ we find a function of $r$, whose minima lie either at $r=0$ for both solutions, or at
\begin{equation}
{\rm a})\, r^3=\frac{9}{2\, L^3\,\mu_4\,V_4}\,P_{\phi}\qquad {\rm b})\,r^3=\frac{9\,n}{4\, L^3\,\mu_4\,V_4}\,P_{\psi}
\end{equation}
respectively. Finally, the on-shell Hamiltonian at these points, for both the $r=0$ and the corresponding $r\ne 0$ solution, is
\begin{equation}
\label{dual_giant_energy}
{\rm a})\, L\,\mathcal{H}=3\,P_{\phi}\qquad {\rm b})\, L\,\mathcal{H}=\frac{3}{2}\,n\,P_{\psi} \,.
\end{equation}

The $r=0$ solutions correspond to a collapsed brane, which looks like a point-like object. 
Consequently we recover the results from the previous section. 
The $r\neq 0$ solutions are the expanded ``dual giant graviton" branes.
They are degenerate both in energy and charges with the point-like solutions, 
and hence they correspond to the same state in the dual field theory
(as we will see below, this is a mesonic operator with no insertion of vector multiplet scalars). 
As usual \cite{Lin:2004nb} we expect the point-like and expanded configurations to have different regimes of validity
in terms of their back-reaction. For a given choice of charges only one type of configuration will lead to a non-singular background.

We would like to stress again that the branes live at $\alpha=0$, 
which is a singular point in the background. Nevertheless their world-volume theory (\ref{D4_action}) is perfectly well-defined.

Finally let us note that we could go back and consider the most generic configuration where we assume $r=r(t)$, $\alpha=\alpha(t)$ and $\theta=\theta(t)$. However one can see that the minimal energy configuration is attained when the corresponding momenta and velocities vanish, thereby recovering 
our original ansatz.

\section{Symplectic quantization} \label{symplecticquantization}

In the previous section we found a dynamical system with a phase space $X$ parametrized by a set of coordinates $Q^A=\{r,\,\alpha,\,\psi,\,\theta,\,\phi\}$  and canonically conjugated momenta $P_A=\{P_r,\,P_{\alpha},\,P_{\psi},\,P_{\theta},\,P_{\phi}\}$. On general grounds, a classical system is defined once we define the symplectic space $(X,\,\omega)$ made out of phase space $X$ and a symplectic structure $\omega$. The quantization of such a system amounts to 
assigning to $X$ a Hilbert space $\mathscr{H}(X,\,\omega)$, where the quantum wave-functions live. 
Following the $AdS_5/CFT_4$ example \cite{Mandal:2006tk, Martelli:2006vh} (see also \cite{Kinney:2005ej, minwalla}), 
by quantizing the phase space of the giant gravitons we should recover the field theory space of dual operators. 

The canonical Poisson brackets are 
\begin{equation}
\{Q^A,\,Q^B\}_{PB}=0\qquad \{P_A,\,P_B\}_{PB}=0\qquad \{Q^A,\,P_B\}_{PB}=\delta^A_{B} \,.
\end{equation}
Let us denote the constraints for the dynamical system for the two types of solutions as $\{f^{{\rm a})}_A,\,f^{{\rm b})}_A\}$.
These are given by $f_r^{(a,b)} =  P_r $, $ f_\alpha^{(a,b)} = P_\alpha$, $f_\theta^{(a,b)} = P_\theta$, and
\begin{equation}
\begin{array}{ll}
f_\psi^{(a)}  =  P_{\psi}-\frac{2}{n}\,\frac{2\,L^3\,V_4\,\mu_4}{9}\,r^3\,\cos\theta  & f_\phi^{(a)} = P_{\phi}-\frac{2\,L^3\,V_4\,\mu_4}{9}\,r^3 \\[5pt]
f_\psi^{(b)} = P_{\psi}-\frac{4\,L^3\,\mu_4\,V_4}{9\,n}\,r^3 & f_\phi^{(b)} = P_{\phi}-\frac{n}{2}\,\frac{4\,L^3\,\mu_4\,V_4}{9\,n}\,r^3\,\cos\theta \,.
\end{array}
\end{equation}
The equations of motion impose the constraints $f_A^{(a,b)}=0$ on the phase space. 
Define the matrices $M^{(a,b)}_{AB}\equiv \{f^{(a,b)}_A,f^{(a,b)}_B\}$.
Since the constraints involving $\alpha$ are trivial, we can eliminate the corresponding row and column from $M^{(a,b)}$,
and reduce $X$ to an eight dimensional space.

The symplectic structure on the reduced phase space is obtained by computing the Dirac bracket, which in this case is
\begin{equation}
\{Q^A,\,Q^B\}_{DB}=(M_{AB})^{-1} \,.
\end{equation}
The symplectic structure for the two solutions is then given by
\begin{equation}
\begin{array}{l}
\omega_{{a}}=\frac{4\,L^3\,\mu_4\,V_4}{3\,n}\,r^2\,\cos\theta\,dr\wedge d\psi+\frac{2}{3}\,L^3\,\mu_4\,V_4\,r^2\,dr\wedge d\phi+\frac{4\,L^3\,\mu_4\,V_4}{9\,n}\,r^2\,\sin\theta\,d\psi\wedge d\theta\\ \\
\omega_{{b}}=\frac{4\,L^3\,\mu_4\,V_4}{3\,n}\,r^2\,dr\wedge d\psi+\frac{2}{3}\,L^3\,\mu_4\,V_4\,r^2\,\cos\theta\,dr\wedge d\phi-\frac{2}{9}\,L^3\,\mu_4\,V_4\,r^3\,\sin\theta\,d\theta\wedge d\phi \,.
\end{array}
\end{equation}
Integrating, we get the one-forms 
\begin{equation}
\begin{array}{l}
\nu_{{a}}=\frac{2\,L^3\,\mu_4\,V_4}{9}\,r^3\,(d\phi+\frac{2}{n}\,\cos\theta\,d\psi) \qquad 
\nu_{{b}}= \frac{2\,L^3\,\mu_4\,V_4}{9}\,r^3\,(\frac{2}{n}\,d\psi+\cos\theta\,d\phi) \,.
\end{array}
\end{equation}
Recall that we have rescaled the $\psi$ coordinate in the original metric so as to have period $2\pi$, while at the same time the giant moves along the $-\phi$ direction. Let us go back to the original coordinates. Besides, let us introduce 
$\rho^2 \equiv (4/9)\mu_4 V_4 L^3 r^3$, so that
\begin{equation}
\label{symplectic_one_forms}
\hat{\nu}_{{a}}=\frac{\rho^2}{2}\,(d\phi-\cos\theta\,d\psi) \qquad \hat{\nu}_{{b}}=\frac{\rho^2}{2}\,(d\psi-\cos\theta\,d\phi) \,.
\end{equation}

Having determined the symplectic form we now have a symplectic manifold $(X,\,\omega)$. We would like now to quantize this system. This amounts to associating to this classical phase space the Hilbert space $\mathscr{H}(X,\,\omega)$ of wave-functions for the quantized system. One would be naturally tempted to simply define as $\mathscr{H}(X,\,\omega)$ the space of functions on $(X,\,\omega)$. However this way wave-functions would generically depend on all coordinates on $(X,\,\omega)$, that is, on both momenta and position. As reviewed in \cite{Martelli:2006vh}, the correct quantization prescription is to identify $\mathscr{H}(X,\,\omega)$ with the space of holomorphic functions, in the complex structure defined by $\omega$, on $(X,\,\omega)$. In this way, wave-functions naturally depend only on half of the coordinates of the phase space. Thus, the upshot is that the Hilbert space associated to the classical system of giant gravitons consists of holomorphic functions on the classical space $(X,\,\omega)$.

In order to understand the classical space $(X,\,\omega)$, in particular with the above $\omega_{{a},\,{b}}$, 
consider an auxiliary $\mathbb{C}^2$ parametrized by $(z_1,\,z_2)$.  The metric, $ds^2=dz_i\,d\bar{z}_i$,
can be rewritten in two equivalent ways:
\begin{equation}
ds^2=\left(\begin{array}{c c} d\bar{z}_1 & d\bar{z}_2 \end{array}\right)\,\left( \begin{array}{c c} 1 & 0 \\ & \\ 0 & 1\end{array}\right)\,\left(\begin{array}{c} dz_1 \\ \\dz_2\end{array}\right) \quad {\rm or}\quad   ds^2=\left(\begin{array}{c c} d\bar{z}_1 & dz_2 \end{array}\right)\,\left( \begin{array}{c c} 1 & 0 \\ & \\ 0 & 1\end{array}\right)\,\left(\begin{array}{c} dz_1 \\ \\d\bar{z}_2\end{array}\right) \,.\nonumber
\end{equation}
This shows that $\mathbb{C}^2$ is invariant under $SU(2)_a\times SU(2)_b$,
where $SU(2)_a$ rotates $(z_1,z_2)$ and $SU(2)_b$ rotates $(z_1,\bar{z}_2)$.
We can define two complex structures on $\mathbb{C}^2$,
\begin{equation}
J_{{a}}=i\,\,(dz_1\,\wedge d\bar{z}_1+dz_2\,\wedge d\bar{z}_2) \qquad J_{{b}}=i\,(dz_1\,\wedge d\bar{z}_1-dz_2\,\wedge d\bar{z}_2) \,.
\end{equation}
The first is invariant under $SU(2)_a\times U(1)_b$, where $U(1)_b$ is the Cartan subgroup of $SU(2)_b$,
and the second is invariant under $SU(2)_b\times U(1)_a$.
Let us express these in polar coordinates:
\begin{equation}
z_1=\rho\,e^{i\frac{\psi+\phi}{2}}\,\sin\frac{\theta}{2}\qquad z_2=\rho\,e^{i\frac{-\psi+\phi}{2}}\,\cos\frac{\theta}{2} \,,
\end{equation}
where $\psi\sim \psi + 4\pi$, $\phi\sim \phi + 2\pi$ and $0\leq\theta\leq \pi$.
The periodic coordinates $\psi$, $\phi$ are shifted by the $U(1)_a$ and $U(1)_b$ Cartan subgroup, respectively.
In these coordinates
\begin{equation}
\begin{array}{l}
J_{{a}}=\rho\,d\rho\wedge d\phi-\rho\,\cos\theta\,d\rho\wedge d\psi+\frac{1}{2}\,\rho^2\,\sin\theta\,d\phi\wedge 
d\psi=d\Big[\frac{\rho^2}{2}\,(d\phi-\cos\theta\,d\psi)\Big] \\  \\
J_{{b}}=-\rho\,\cos\theta\,d\rho\wedge d\phi+\rho\,d\rho\wedge d\psi+\frac{1}{2}\,\rho^2\,\sin\theta\,d\theta\wedge d\phi
=d\Big[\frac{\rho^2}{2}\,(d\psi-\cos\theta\,d\phi)\Big] \,.
\end{array}
\end{equation}
Now consider the orbifold $\mathbb{C}^2/\mathbb{Z}_n$ where $\mathbb{Z}_n$ acts as 
\begin{equation}
(z_1,\,z_2)\, \rightarrow \, (\omega\,z_1,\,\omega^{-1}\,z_2)\,\qquad \omega^n=1 \,.
\end{equation}
This breaks $SU(2)_a \rightarrow U(1)_a$ (for $n>2$) and preserves $SU(2)_b$.
In polar coordinates it simply changes the periodicity of $\psi$ to $\psi\sim \psi + 4\pi/n$.
On the orbifold, the first complex structure $J_a$ preserves $U(1)_a\times U(1)_b$,
whereas the second complex structure $J_b$ preserves $U(1)_a\times SU(2)_b$.

Comparing with the symplectic one-forms (\ref{symplectic_one_forms}) we see that 
$J_{{a},\,{b}}=d\,\hat{\nu}_{{a},\,{b}}$.
The geometric quantization of the phase space of dual giant gravitons is therefore mapped to that of 
$\mathbb{C}^2/\mathbb{Z}_n$.
The wave-functions correspond to holomorphic functions on $\mathbb{C}^2/\mathbb{Z}_n$
with a given complex structure, $J_a$ or $J_b$, depending on whether $P_\phi/P_\psi$
is larger or smaller than $n/2$,
and are classified according to the corresponding symmetry, $U(1)_a\times U(1)_b$
or $U(1)_a\times SU(2)_b$, respectively.
Therefore there is a one-to-one map between wave-functions on $\mathbb{C}^2/\mathbb{Z}_2$, 
the geometrically quantized phase space of dual giants,
and mesonic operators in the field theory. 
In fact, the translation to the field theory language is now obvious: $SU(2)_{{b}}$ corresponds to the $SU(2)_R$ R-symmetry, and 
$U(1)_{{a}}\in SU(2)_a$ corresponds to the $U(1)_M\in SU(2)_M$ mesonic symmetry. 

Thus, although they live in the near-horizon $AdS_6\times S^4/\mathbb{Z}_n$ space,
the dual giant gravitons located at $\alpha=0$ actually probe the 
$\mathbb{C}^2/\mathbb{Z}_n$ space transverse to the D4-branes inside the O8-plane,
which is the Higgs branch of the theory.

\section{Field theory operators}\label{fieldtheory}

The dual giant gravitons should correspond to a sub-sector of operators on the Higgs branch that are flavor-, baryon- and 
instanton-neutral.
These operators involve only the bi-fundamental and antisymmetric hyper-multiplets, and 
are classified 
by their quantum numbers under $SU(2)_R\times U(1)_M$. 

Since the sub-sector we are interested in only involves hyper-multiplets, it turns out to be technically easier to consider the field theory on 
$\mathbb{R}^{1,\,3}\times S^1$. Upon sending the radius of the $S^1$ to zero we find a 4d theory whose quiver diagram and interactions 
are precisely equal to those of the original theory. From the 4d point of view, it is natural to choose an 
$\mathcal{N}=1$ sub-algebra and express the theory in terms of $\mathcal{N}=1$ super-fields. The natural object to consider then is the chiral ring, composed of chiral operators upon imposing the equivalence relations dictated by the F-terms. Note that in 4d our theories really have $\mathcal{N}=2$ supersymmetry, 
where the R-symmetry is $SU(2)_R\times U(1)'_R$. 
The $SU(2)_R$ part is inherited from the 5d R-symmetry, and the $U(1)'_R$ part arises from the compactification. 
However in the ${\cal N}=1$ chiral ring only the Cartan $U(1)_R\in SU(2)_R$ is manifest.
For example, a hyper-multiplet corresponds to a pair of 
chiral super-fields $(Q,\,\tilde{Q})$ in conjugate representations of the gauge group, whereas
$SU(2)_R$ acts on the doublet $(Q,\,\tilde{Q}^{\dagger})$, \textit{i.e.} in a non-holomorphic way. 
The chiral ring therefore automatically chooses the complex structure
$J_a$, and will therefore only include the subset of operators 
that are dual to the dual giant graviton states
corresponding to this choice of complex structure. Note that the other subset of giants just corresponds to non-holomorphic operators in this language. For this reason in the following we will concentrate on those operators/giants which are holomorphic in the chosen $\mathcal{N}=1$ language.

Let us start with the $n=1$ case. This is a $USp(2\,N)$ theory with one antisymmetric hyper-multiplet $A$.
The fundamental hyper-multiplets do not play a role in the sector in question, so we set $N_f=0$.
In the 4d ${\cal N}=1$ language $A$ corresponds to a pair of antisymmetric chiral superfields  $(A_1,A_2)$,
transforming as a doublet under $SU(2)_M$.
The interactions are captured by the super-potential \cite{Benvenuti:2010pq}
\begin{equation}
W=\epsilon^{\alpha\beta}{\rm Tr}\, (A_{\alpha}\Phi A_{\beta}) \,,
\end{equation}
%
where $\Phi$ is the adjoint chiral super-field of the ${\cal N}=2$ vector multiplet.
The components $(A_1,A_2)$ carry charges $(1/2, -1/2)$, respectively,
under $U(1)_M\in SU(2)_M$, and charges $(1/2,1/2)$ under $U(1)_R\in SU(2)_R$.
The F-term is given by
\begin{equation}
 \epsilon^{\alpha\beta}\,A_{\beta}\,A_{\alpha}=0 \,.
\end{equation}
The effect of this F-term is to symmetrize products of $A_{\alpha}$. 
All of the operators in question can therefore be expressed as
\begin{equation}
\mathcal{O}_{m,\,n}={\rm Tr}\,(A_1^m\,A_2^n) \,.
\end{equation}
Having constructed the operators by reduction to 4d, we need to come back to 5d. In 5d these operators have 
$\Delta=\frac{3}{2}\,(n+m)$ and $Q_{{R}}=\frac{1}{2}(n+m)$, so that they satisfy 
\begin{equation}
\Delta=3\,Q_{{R}} \,.
\end{equation}
Upon identifying $Q_R$ with $P_\phi$ this
agrees with the energy of the corresponding dual giant graviton (\ref{dual_giant_energy}),
as it should in global $AdS$.
The $U(1)_M$ charge of these operators is $Q_M=\frac{1}{2}(n-m)$. 
We see that $|Q_M| \leq Q_R$.
Identifying $Q_M$ with $\frac{n}{2}P_\psi$, this agrees with
the condition for the dual giant graviton (\ref{choices})
(recall that the $\psi$ coordinate in (\ref{choices}) was rescaled so that $\psi\sim \psi + 2\pi$).
For the first few operators we find
\begin{equation}
\begin{array}{l |  l |  l}
\frac{2}{3}\,\Delta & {\rm operators} & \# \\ \hline
1 & A_1,\,A_2 & t\,(z+z^{-1}) \\
2 & A_1^2,\,A_1\,A_2,\,A_2^2 & t^2\,(z^2+1+z^{-2}) \\
3& A_1^3,\,A_1^2\,A_2,\,A_1\,A_2^2,\,A_2^3 & t^3\,(z^3+z^1+z^{-1}+z^{-3}) \\
\end{array}
\end{equation}
where we introduced the fugacity $t$ which stands for the dimension of the operator and the fugacity $z$ which counts the $Q_M$ charge. Note that $z$ appears through the character of the highest weight $m$ $SU(2)$ representation, which we will denote as $[m]_z$. It is straightforward to see that the generating function is given by
\begin{equation}
\sum_{m} [m]_z\,t^m=\frac{1}{(1-t\,z)\,(1-\frac{t}{z})} \,.
\end{equation}
This is precisely the Hilbert series of $\mathbb{C}^2$, meaning that these operators are in one-to-one correspondence with holomorphic functions on $\mathbb{C}^2$. This is precisely the expected result for the case $n=1$.

Let us now consider the case of $n=2$, focusing first on the NVS case with the gauge group $SU(2N)$,
and two antisymmetric hyper-multiplets $A,\, A'$.
In terms of the pairs of chiral super-fields $(A_1,A_2)$ and $(A'_1,A'_2)$,
the 4d super-potential is given by
\begin{equation}
W=\epsilon^{\alpha\beta} {\rm Tr}\, (A_\alpha\,\Phi\, A_\beta +A'_\alpha\,\Phi\, A'_\beta) \,.
\end{equation}
%
In this case the global $SU(2)_M$ acts on the doublets $(A_1,A'_1)$ and $(A'_2,A_2)$.
In particular, $A_1$ and $A'_2$ carry a $U(1)_M$ charge of $+1/2$, and $A'_1$ and $A_2$ carry a $U(1)_M$
charge of $-1/2$.
The $SU(2)_R$ symmetry acts on $(A_1,A_2^\dagger)$ and $(A_1',A_2^{'\dagger})$,
so the $U(1)_R$ charge assignment is $+1/2$ for all $A_\alpha,\, A'_\alpha$.\footnote{Note that there is one more symmetry 
assigning charge 
$1/2$ to the $A_\alpha$ and $-1/2$ to the $\tilde{A}_\alpha$. However no mesonic operators is charged under this symmetry, which is thus a baryonic $U(1)$. In this paper we are not concerned about baryonic operators.}
Taking into account the F-term, which imposes $\epsilon^{\alpha\beta}(A_\beta A_\alpha + A'_\beta A'_\alpha)=0$, 
the first few operators are given as follows
\begin{equation}
\begin{array}{l |  l |  l}
\frac{2}{3}\,\Delta & {\rm operators} & \# \\ \hline
2 & \,A_1 A'_2,\, A_1 A_2,\, A'_1 A_2 & t^2\,(z^2+1+z^{-2}) \\[2pt]
4 & A_1 A'_2 A_1 A'_2,\, A_1 A_2 A_1 A'_2,\, A_1 A_2 A_1 A_2, 
& t^4\,(z^4+z^2+1+z^{-2}+z^{-4}) \\[2pt]
 & \,A_1 A_2 A'_1 A_2,\, A'_1 A_2 A'_1 A_2 & 
\end{array} \,.
\end{equation}
These satisfy $\Delta=3\,Q_{R}$ and $|Q_M|\leq Q_{{R}}$, which are again the expected relationships
for the dual giant gravitons. 
We also recognize here the first few terms in the expansion of
\begin{equation}
\frac{(1-t^4)}{(1-t^2)\,(1-t^2\,b^2)\,(1-\frac{t^2}{b^2})} \,,
\end{equation}
which is the Hilbert series for $\mathbb{C}^2/\mathbb{Z}_2$, thus precisely recovering the dual giant graviton result.

In the VS case we have a $USp(2\,N)\times USp(2N)$ theory with one bi-fundamental hyper-multiplet,
which we express in terms of 4d chiral super-fields as $(Q,\tilde{Q})$. 
The $U(1)_M$ charge assignment is $(1/2,-1/2)$, and the $U(1)_R$ charge assignment is $(1/2,1/2)$.
The F-term imposes $Q\,\tilde{Q}=\tilde{Q}\,Q$. The first few operators are given by 
\begin{equation}
\begin{array}{l |  l |  l}
\frac{2}{3}\,\Delta & {\rm operators} & \# \\ \hline
2 & \,Q^2,\,Q\,\tilde{Q},\,\tilde{Q}^2 & t^2\,(z^2+1+z^{-2}) \\
4 & Q^4,\,Q^3\,\tilde{Q},\,Q^2\,\tilde{Q}^2,\,Q\,\tilde{Q}^3,\,\tilde{Q}^4 & t^4\,(z^4+z^2+1+z^{-2}+z^{-4}) \,.
\end{array}
\end{equation}
We again see that  $\Delta=3\,Q_{{R}}$ and $|Q_M|\leq Q_R$, as expected, and 
we again find the Hilbert series for $\mathbb{C}^2/\mathbb{Z}_2$, as expected from the giant graviton analysis.

Note that the analysis above is strictly speaking valid only at large $N$, as we have neglected possible relations among traces. However, a complete analysis at finite $N$ can be performed explicitly for some small values of $N$ and $n=1,2$
by computing the exact Hilbert series on the Higgs branch as arising from the field theory with the help of the algebraic-geometry symbolic computation program \verb+Macaulay2+ \cite{Macaulay2}. We find that the Hilbert series reproduces the expected $\mathbb{C}^2/\mathbb{Z}_n$ for $n=1,2$.

While we leave a more thorough analysis of the general case for future work, all in all, based on the examples, we expect that the counting of operators in the zero baryonic charge, zero instanton charge and zero flavor charge sector matches exactly the quantization of the phase space of giant gravitons. More explicitly, the operators in this zero-charges sector of the Higgs branch are expected to be in one-to-one correspondence with holomorphic functions on $\mathbb{C}^2/\mathbb{Z}_n$.\footnote{Indeed, the same situation is found in the more familiar $AdS_5/CFT_4$ case for $A_n$ quivers \cite{DHRG}.}

It is interesting to revisit now the status of the gravity computation, where we found two degenerate solutions for each choice of quantum numbers, 
namely the expanded and singular configurations. 
As we have just argued, the dual operator is a meson composed of hyper-multiplets without vector multiplet scalars. As usual, a short meson whose dimension is $\mathcal{O}(1)$ corresponds to a SUGRA fluctuation, \textit{i.e.} point-like particles following BPS geodesics. On the other hand, as the dimension increases, by the time we consider a long meson whose dimension is $\mathcal{O}(N)$, the dual configuration is best described as the expanded brane configuration. Indeed, we expect that if we were to consider the fully back-reacted geometry, the non-singular geometry corresponding to $\mathcal{O}(N)$ is that arising from the back-reaction of the expanded brane configuration, pretty much as in the LLM case \cite{Lin:2004nb}. 
For the purpose of counting operators however, we could just consider the expanded configurations.

\section{Conclusions}\label{conclusions}

In this paper we have studied the sub-sector of the Higgs branch which is both flavor and instanton blind. In terms of the fields in the corresponding quiver theories, it consists of the operators made only out of bi-fundamental and/or antisymmetric hyper-multiplets, with strictly zero baryonic and instantonic charges. In the gravity dual such operators can be put in correspondence with dual giant gravitons, namely D4-branes in global $AdS_6$,
which follow massless geodesics in the internal space. The geometric quantization of the phase space associated to such branes shows that the corresponding operators are in one-to-one correspondence with holomorphic functions on $\mathbb{C}^2/\mathbb{Z}_n$. In fact, we can think of this space as that transverse to the D4-branes inside the O8-plane in the pre-near-horizon background. 
Conversely, at least for the simplest examples, we recover the same results from the field theory perspective. It would be interesting to check this result more thoroughly for the three  whole families. We took a somewhat lengthier route in that we reduced the theory down to 4d, in order to use the more familiar $\mathcal{N}=1$ superspace. It would be interesting to overcome this technicality by working directly in 5d.

The partition function which counts the operators in question corresponds to the Hilbert series of the orbifold. It is natural to expect that this corresponds to the Hilbert series of the entire Higgs branch upon setting to zero the flavor and instanton fugacities. It would be certainly very interesting to go beyond this flavor and instanton blind sector. Besides, it would be interesting to clarify whether this Hilbert series can be thought of as a limit of the super-conformal index \cite{Kim:2012gu}, in the spirit of the corresponding relation for 4d theories found in  \cite{Gadde:2011uv}. 

Having identified the dual giant gravitons, it is natural to wonder whether genuine giant gravitons, namely those expanding in the internal part of the geometry, exist. 
We expect these to correspond to anti-symmetrized products of fields.
In particular, there is an upper limit on the number 
of fields corresponding to the maximal giant graviton, which is a manifestation 
of the so-called string exclusion principle (see \cite{Balasubramanian:2001nh, Corley:2001zk} for the description of such phenomenon in the $AdS_5/CFT_4$ case). Taking for definiteness the $n=1$ case, the natural candidate for the maximal giant would be the Pfaffian operator ${\rm Pf}(A)$. However, as discussed in \cite{Bergman:2012kr}, this operator is related to the $N$-th power of the meson. While this naively suggests that in this case giant gravitons will be absent, certainly a more thorough analysis should be performed. Furthermore, it is natural to ask wether a microscopic description along the lines of   \cite{Janssen:2002cf,Janssen:2003ri,Janssen:2004jz,Janssen:2004cd} is possible. We leave such questions open for future investigations.

\section*{Acknowledgements}

D.R-G. thanks Rak-Kyeong Seong for useful conversations and computing help with \verb+Macaulay2+. He also thanks the Korea Institute for Advanced Study for warm hospitality while this work was in progress. D.R-G. is supported by the Aly Kaufman fellowship. He also acknowledges partial support from the Israel Science Foundation under grant no 
392/09 and from the Spanish Ministry of Science through the research grant FPA2009-07122 and Spanish Consolider-Ingenio 2010 Programme CPAN (CSD2007-00042). O.B. is supported in part by the Israel Science Foundation under grant no. 392/09,
and the US-Israel Binational Science Foundation under grant no. 2008-072.

\end{document}